\documentclass{pasa}%

\usepackage{graphicx}

\usepackage{amsmath}
\usepackage{rotating}

\title[The stellar rotation-activity relation for a sample of SuperWASP and ASAS-SN field stars]{The stellar rotation-activity relation for a sample of SuperWASP and ASAS-SN field stars}

\author[Thiemann et al.]{Heidi B. Thiemann$^1{}^2$, Andrew J. Norton$^1$ and Ulrich C. Kolb$^1$
\affil{$^1$School of Physical Sciences, The Open University, Walton Hall, Milton Keynes, MK7 6AA, UK}%
\affil{$^2$DISCnet Centre for Doctoral Training, The Open University, Walton Hall, Milton Keynes, MK7 6AA, UK}
}%

\jid{PASA}
\doi{10.1017/pas.\the\year.xxx}
\jyear{\the\year}

\usepackage{aas_macros}
\usepackage{hyperref} 
\hypersetup{colorlinks,citecolor=blue,linkcolor=blue,urlcolor=blue}

\begin{document}

\begin{frontmatter}
\maketitle

\begin{abstract}
It is well established that late-type main-sequence stars display a relationship between X-ray activity and the Rossby number, $Ro$, the ratio of rotation period to the convective turnover time. This manifests itself as a saturated regime (where X-ray activity is constant) and an unsaturated regime (where X-ray activity anti-correlates with the Rossby number). However, this relationship breaks down for the fastest rotators. We cross-correlated SuperWASP visually classified photometric light curves and ASAS-SN automatically classified photometric light curves with XMM-Newton X-ray observations to identify 3,178 stars displaying a photometrically defined rotational modulation in their light curve and corresponding X-ray observations. We fitted a power-law to characterise the rotation-activity relation of 900 main-sequence stars. We identified that automatically classified rotationally modulated light curves are not as reliable as visually classified light curves for this work. We found a power-law index in the unsaturated regime of G- to M-type stars of $\beta=-1.84\pm0.18$ for the SuperWASP catalogue, in line with the canonical value of $\beta=-2$. We find evidence of supersaturation in the fastest rotating K-type stars, with a power-law index of $\beta_{s}=1.42\pm0.26$.
\end{abstract}

\end{frontmatter}

%__________________________________________________________________

\section{INTRODUCTION }
\label{sec:intro}

Every star has a magnetic field, and an important product of such a field is X-ray emission. The majority of X-ray emissions originate from active regions - concentrated areas of magnetic fields on the stellar surface such as star spots, collisions between the wind and circumstellar material \citep{parkin}, and small-scale wind shocks \citep{Lucy1980}.

In solar and late-type stars, the stellar magnetic field is generated in a magnetically confined plasma \citep{Vaiana1981}. This plasma is driven by the stellar magnetic dynamo, which is in turn driven by the internal stellar differential rotation (\citet{Parker1955}; \citet{Wilson1966}; \citet{Kraft1967}). This process was confirmed in the Sun through helioseismology \citep{Duvall1984}.

Over the lifetime of a star, the rotation rate changes significantly. As a star ages and moves from the Pre-Main Sequence (PMS) to the Main Sequence (MS), it experiences spin-down, the internal dynamo weakens, and as a result, the coronal X-ray emission decreases \citep{wright2011}. Magnetic braking is active in PMS phase, but opposed by collapse/contraction which would speed up the star, and continues on the MS. Spin-down is driven primarily by mass loss through the magnetised stellar winds steadily transferring angular momentum away from the star. For MS stars, this decline in rotation can be approximated by Skumanich's law, $\Omega_e \propto t^{-1/2}$ \citep{Skumanich1972}, where $\Omega_e$ is the angular velocity of the star's equator, and $t$ is the age of the star. 

Following this work, a relationship between rotation and activity was quantified by \citet{Pallavicini1981}, who found that X-ray luminosity scaled as $L_X \propto (v\sin i)^{1.9}$, providing the first evidence for the dynamo-induced nature of stellar coronal activity. Subsequently, \citet{Noyes1984} combined these parameters into the Rossby number, $Ro = P_{rot}/\tau$, as the ratio of stellar rotation period, $P_{rot}$, to the convective turnover time, $\tau$, of plasma covering a certain distance in the convective envelope, and which acts as a parameter for the stellar dynamo \citep{Landin2010}.

Numerous studies of late-type MS stars (e.g., \citet{Pizzolato2003}, \citet{wright2011}) have characterised the relationship between stellar activity and rotation period, dividing the rotation activity into two main regimes for rotation periods of $\sim 1 - 10d$: the unsaturated and saturated regimes, and a third possible regime, the supersaturated regime.

In the unsaturated regime, for $Ro > 0.13$, the fractional X-ray luminosity, $L_x/L_{bol}$, anticorrelates with $Ro$. \citet{wright2011} finds a power-law slope of $\beta \sim -2.7$ for F- to M- type MS stars. This strong rotation-activity relation indicates that the dynamo is responsible for stellar activity levels. 

For fast rotating solar and late-type MS stars, the relationship is found to break down \citep{Micela1985}, and saturation of $L_x/L_{bol}$ occurs at $Ro \sim 0.13$, regardless of spectral type \citep{Vilhu1984}, with $L_x/L_{bol}$ saturating at $L_x/L_{bol} \sim 10^{-3}$. At this point X-ray emission becomes a function of bolometric luminosity alone \citep{Pizzolato2003}, i.e. the colour, mass, or radius of the star. The saturation threshold, $Ro_{sat}$, scales with the convective turnover time, $\tau$, such that $Ro$ is constant. The cause of saturation is still debated. Theories include centrifugal stripping of the corona at high rotation rates \citep{Jardine1999}, saturation of the filling factor of active regions on the stellar surface, or a saturation of the dynamo itself \citep{Vilhu1984}.

Studies have found that nearly all fully convective, fast-rotating solar and late-type stars exhibit saturation of coronal X-rays \citep{wright2011}. The rotation period at which saturation occurs is longer for later spectral type stars, due to larger convective turnover times (\citet{Landin2010}, \citet{wright2011}). Whilst surprising, it has also been found that fully convective slow rotators show a solar-type rotation-activity relation \citep{wright2018}. Despite this, a comprehensive and satisfactory dynamo theory that explains both the saturated and unsaturated regimes is yet to be found (\citet{weiss2005}, \citet{Brandenburg2011}).

In addition to these two regimes, studies have indicated that a third regime, supersaturation, may exist for the very fastest rotating MS stars. Supersaturation has been detected in PMS stars, M-dwarfs, and W UMa contact binaries (eg., \citet{Randich1998}, \citet{Argiroffi2016}, \citet{Jeffries2011}, \citet{stepien2001}). In this regime, the X-ray to bolometric luminosity ratio has been observed to decrease slightly below the saturation level (\citet{Jardine1999}, \citet{Jardine2004}, \citet{Jeffries2011}). Although very scarcely observed, supersaturation has been seen to occur in young ($<$ 100 Myr) G- and K-type stars in young clusters \citep{Randich1996}. Whilst \citet{Argiroffi2016} has found evidence for supersaturation in pre-main-sequence (PMS) stars, studies such as \citet{wright2011} have found no statistically significant evidence for supersaturation in MS stars, or any decline in luminosity ratio for fast rotators. 

In this paper, we update the research by \citet{norton2007} with a new cross-correlation of the SuperWASP and XMM-Newton 4XMM-DR9 catalogues, which has detected 16,827 X-ray visible unique objects displaying variability, including 1,257 rotationally modulated variables, and 2,131 binaries and pulsators. Additionally, the ASAS-SN Catalogue of Variable Stars has been cross-correlated against the XMM-Newton 4XMM-DR9 catalogue. These catalogues have been cross-correlated with Gaia-DR2, which contains the parallaxes and proper motions, as well as light curves for $> 550,000$ variable sources.  

%__________________________________________________________________
 
\section{DATA REDUCTION}

The catalogue was primarily compiled from the SuperWASP archive, ASAS-SN Catalogue of Variable Stars, 4XMM-DR9 catalogue, and Gaia-DR2 catalogue, selecting only unique objects displaying a rotational modulation with photometric, X-ray, and parallax data. Here we describe the catalogues used in cross-correlation, the data reduction, and analysis.

%__________________________________________________________________

\subsection{SuperWASP photometric catalogue}

The Wide Angle Search for Planets, SuperWASP, is the most successful ground-based search for transiting exoplanets \citep{Pollacco2006}, having discovered $\sim$200 hot Jupiters since 2004. The entire sky has been surveyed, excluding the Galactic Plane where individual objects cannot be resolved due to the large pixel scale of the telescope. Not only has it discovered numerous exoplanets, but it has also produced an archive of more than 30 million light curves.  

SuperWASP is made up of two identical robotic observatories and a consortium of eight academic institutions. SuperWASP-North is located on the island of La Palma in the Canaries, and SuperWASP-South is situated with the South African Astronomical Observatory (SAAO), in South Africa. Each observatory consists of 8 wide-angle cameras giving each camera a 7.8 x 7.8 square degrees field of view (FoV), giving a total of 500 square degrees of sky coverage. The large pixel size of 16.7 arcsec per pixel means that whilst all-sky coverage can be completed, SuperWASP avoided observing the Galactic Plane as it could not resolve individual stars within this dense area of sky \citep{Pollacco2006}. SuperWASP data provides a high cadence, long baseline all-sky archive, away from the Galactic Plane, for stars between about $V=8-15$ to investigate stellar variability. 

\begin{figure}
   \centering
   \includegraphics[width=\hsize]{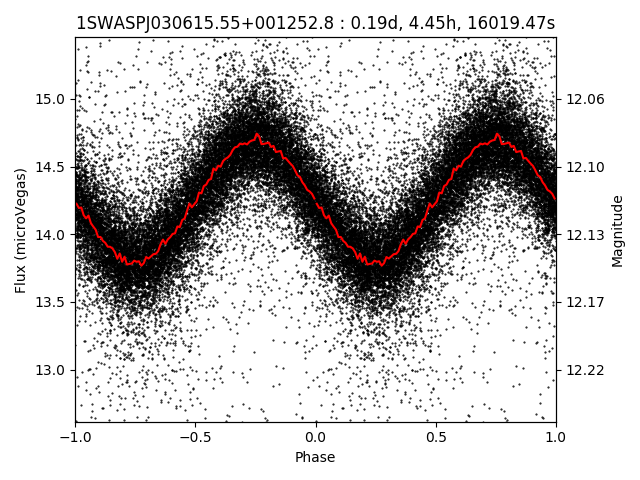}
      \caption{Example of a light curve of a rotator with a period of 16,019 seconds ($\sim$0.19 days) in the SuperWASP archive.}
         \label{wasp}
\end{figure}

\begin{figure}
   \centering
   \includegraphics[width=\hsize]{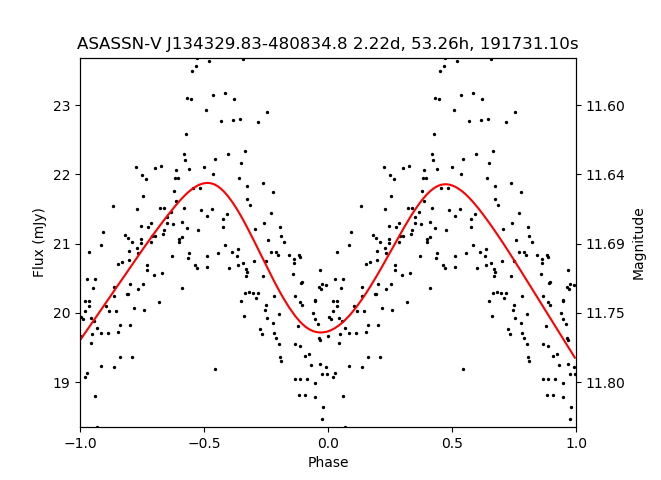}
      \caption{Example of a light curve of a rotator with a period of 191,731 seconds ($\sim$2.22 days) in the ASAS-SN variable stars database.}
         \label{ASAS}
\end{figure}

%__________________________________________________________________

The \textit{SuperWASP periodicity} catalogue is the result of a new search by \citet{Norton2018} for periodic variability in the SuperWASP archive of $\sim$ 31 million light curves. The period searching routine used CLEANed power spectrum analysis and a phase dispersion minimisation folding analysis. This routine detected $\sim$8 million potential periods in $\sim$3 million objects, with a sensitivity of between $\sim$1 hr and $\sim$1 yr. However, many systematic effects occurred in the period search of SuperWASP lightcurves, resulting in the detection of periods of a lunar month, sidereal day, or integer multiples, and fractions of a day (i.e. 1/2, 1/3, 1/4, etc). As a result of these systematic effects, any object "flagged" as having a period close to these values was rejected from this work. All "unflagged" objects are the subject of a citizen science Zooniverse project aiming to classify 1,569,061 periods in 767,199 objects, resulting in the construction of catalogues of types of stellar variable \citep{Norton2018}. The periods in this catalogue have an uncertainty of $\sim\pm0.1\%$.

The \textit{SuperWASP periodicity} catalogue also contains duplicated objects, arising from the detection of multiple periods for a single object. The most likely period for each object was found through identifying the most significant period, and any duplicates were removed. Any remaining duplicated objects were removed by performing an internal positional cross-correlation, removing objects within a 2$^{\prime}$ radius of each other ($>$ 4 times the photometric aperture radius, 34$^{\prime \prime}$). 

The remaining 767,199 ($\sim$2.5\% of objects in SuperWASP archive) genuine objects with detectable periodic variability on timescales from hours to years are ready to be cross correlated with X-ray data. Figure \ref{wasp} gives an example of the light curve of a rotationally modulated star in the SuperWASP archive.

%__________________________________________________________________

\subsection{ASAS-SN photometric catalogue}

The All-Sky Automated Survey for Supernovae (ASAS-SN) is an optical survey which has monitored the entire sky since 2014 \citep{2019asas}. ASAS-SN is currently composed of 24 telescopes based in Hawaii, Chile, Texas, South Africa, and China. Each camera has a field of view of 4.5deg$^2$, and a pixel scale of 8$^{\prime \prime}$. Despite the relatively large pixel scale, the typical astrometric error is $\sim$1$^{\prime \prime}$. ASAS-SN uses APASS \citep{apass} magnitudes which tend to deviate from the V-band ASAS-SN magnitudes for V$>$14 due to blended sources where multiple APASS sources are detected by a single ASAS-SN pixel.

Like SuperWASP, this survey did not aim to study stellar variability, instead focusing on bright supernovae, but given the cadence of $\sim$2-3 days, and a limiting magnitude of V$\leq$17 mag, ASAS-SN has been able to study $>$ 50 million bright sources. Light curves of variable stars are classified using a random forest classifier, and verified against other catalogues, as described in \cite{2018asasJ}. Cross-correlated with other cataglogues, e.g. NOMAD, 99\% of ASAS-SN variables detected are within 5.2$^{\prime \prime}$ of a previously catalogued star.  The ASAS-SN Catalogue of Variable Stars contains 666,502 light curves, of which 90,712 belong to rotational variables. Each object in the ASAS-SN Catalogue of Variable Stars contains information about the variability type, classification probability, period, and magnitude, amongst other values. For this work, we selected stars with a classification probability of $>$0.75.

Figure \ref{ASAS} gives an example of the light curve of a rotationally modulated star in the ASAS-SN Catalogue of Variable Stars.

%__________________________________________________________________

\subsection{XMM-Newton X-ray catalogue}

The 4XMM-DR9 catalogue contains detections of 550,124 unique X-ray sources from 11,204 observations made by the XMM-Newton EPIC cameras covering 1089 square degrees. The detections cover the energy range 0.2 - 12 keV and span 19 years between 3 February 2000 and 26 February 2019. This work uses the broad energy band (0.2 - 4.5 keV).

Low-mass stellar X-ray sources are known to show high levels of variability due to magnetic flares \citep{Caramazza2007} and rotational modulation. Bright flares can significantly increase the X-ray flux from a source over the course of a short observation. Short-lived increases in X-ray flux should be removed through multiple observations of the same source, however the majority ($\sim$80\%) of objects in the 4XMM-DR9 catalogue contain only one detection of the source, with only $\sim$41\% of these objects having an uncertainty in the soft X-ray band of $\leq 20\%$ in flux. $\sim$71\% have an uncertainty of $\leq 30\%$ in flux. We expect this to be one of the most significant sources of uncertainty in our work.

As with the \textit{SuperWASP periodicity} catalogue, possible duplicated objects were removed by performing an internal positional cross-correlation, removing objects within a 24$^{\prime \prime}$ radius of each other (4 times the uncertainty in the EPIC extent radius). The remaining objects are assumed to be unique.

The 4XMM-DR9 catalogue contains a summary flag for each source, from 0-11. For example, a flag of 0 indicates a 'good' source, a flag of 1 indicates that parameters may be affected, a flag of 2 indicates a possibly spurious source, and a flag of 3 indicates that the source is located in an area where there may be a spurious detection. 95.7\% of the sources in our catalogue have a flag$\leq$1. All unique sources with a flag $\geq$2 are removed from the catalogue.

%__________________________________________________________________

\subsection{Additional tables}

For the calculation of bolometric luminosity, photometric colour and parallax are required. The majority of previous studies of the rotation-activity relation have used either the $B - V$ or $V - K_s$ colour as various proxies, i.e. for effective temperature. B, V, R, J, H, K photometric magnitudes were taken from the Two Micron All Sky Survey (2MASS) \citep{2mass} and NOMAD \citep{nomad} catalogues, supplemented by USNOB1.0 \citep{usno}, APASS \citep{apass}, and GCS2.3 \citep{gsc}, and used to derive $V - K_s$ for all stars. $V - K_s$ was used as a proxy for effective temperature and used to derive bolometric corrections. For each star, estimated spectral types were based on \cite{Mamajek}.

The conversion of observed to absolute magnitude requires stellar distance. Due to well known problems with the use of non-corrected Gaia-DR2 parallaxes for distance calculation, the \citet{bailer2018} catalogue was used, which provides corrected distances. A 10$^{\prime \prime}$ cross-correlation was used between the \citet{bailer2018} catalogue and \textit{SuperWASP periodicity} catalogue and ASAS-SN Catalogue of Variable Stars. Both the bolometric luminosity and the X-ray luminosity scale with $d^2$, hence the distance value cancels in the bolometric flux ratio. In this work, distances are instead used to identify main sequence stars and correct for interstellar reddening.

\begin{figure}
   \centering
   \includegraphics[width=\hsize]{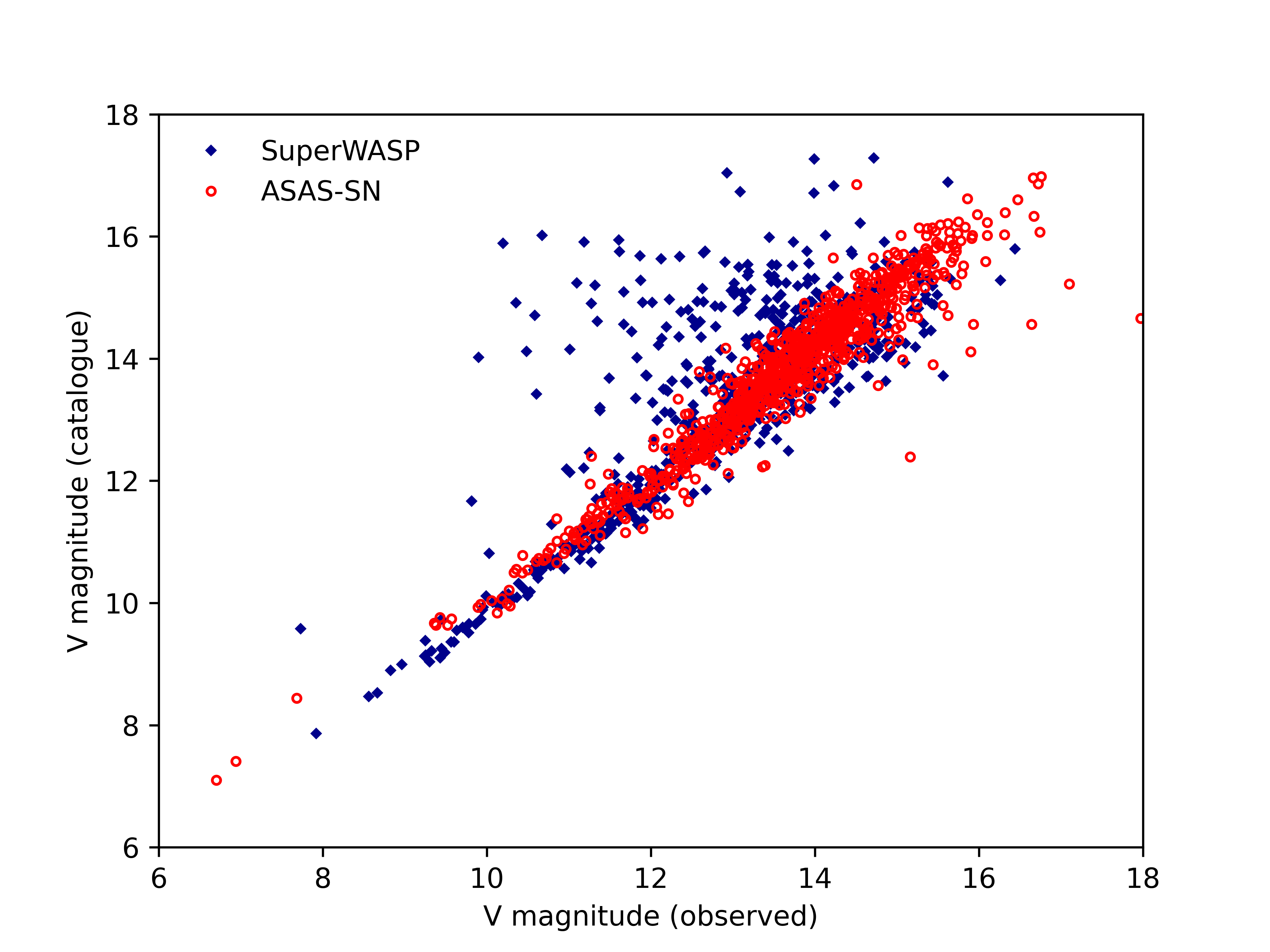}
      \caption{V magnitude (observed) against V magnitude (catalogue). Stars with a V magnitude difference of greater than 1$\sigma$ are removed from the catalogue.}
         \label{Vdiff}
\end{figure}

Such a spatial cross-correlation is likely to pick up non-stellar and non-rotational stellar objects in the process, hence the final cross-correlation is a 10$^{\prime \prime}$ search of SIMBAD tables to identify such objects. Non-stellar and non-rotational stellar objects were removed based on the SIMBAD object type. The catalogue V magnitudes and the observed V magnitudes of the remaining objects were plotted, as shown in Figure \ref{Vdiff}, and objects with a difference in V magnitude of $>$0.85 (or 1 $\sigma$) were removed.

%__________________________________________________________________

\subsection{Cross-correlation}

Through a 2$^{\prime}$ cross-correlation of unique objects in both \textit{SuperWASP periodicity} and 4XMM-DR9 objects using TOPCAT, 39,367 possibly coincident objects with both periodic photometric variability and an X-ray flux were identified. The large cross-correlation radius of 2$^{\prime}$ between the SuperWASP and 4XMM-DR9 catalogues is due to the large (34$^{\prime \prime}$) photometric aperture radius of SuperWASP, and accounts for $\sim$4 times this radius.

Of these, 16,827 were flagged as not having a period close to 1 day. Phase-folded light curves were plotted for 16,827 objects and were visually classified by the authors. 12,509 were rejected as not convincingly variable, 1,232 displayed rotational modulation, and 2,131 were categorised as binaries and pulsators. Of the 1,257 objects, 990 had corresponding colours, and positive, real Gaia-DR2 parallaxes in the \citet{bailer2018} catalogue. Of these 990 objects, 106 were found to be non-stellar or non-rotational stellar objects, and 303 were found to be giant or supergiant stars. Such stars do not follow a rotation-activity relation and are known to experience a sharp decrease in X-ray emission at spectral type K1. Giant and supergiant stars were identified using stellar luminosity by spectral class from \citet{allen2000}. Removing giant stars from the catalogue leaves 581 SuperWASP stars.

The ASAS-SN rotational variable stars were cross-correlated with 4XMM-DR9 with a 10$^{\prime \prime}$ cross-correlation radius, giving 1,921 objects. Of these, 75 appeared in the SuperWASP catalogue and were removed as duplicates. For objects which occurred in both catalogues with a period detection, 72\% of the objects had a period difference of $<$2\%, and 80\% agreed to within $<$10\%. For objects with periods which differed by $>$2\%, each was assessed by eye to determine which was the most likely true period. Of the 1,846 remaining objects, 882 contain information on both period and colour. Of these, 786 have an ASAS-SN classification probability of $>$ 0.75. The ASAS-SN Catalogue of Variable Stars has already been cross-correlated with catalogues of variable stars (e.g. VSX and GCVS) and as such, all 786 objects are main sequence stars. 115 SuperWASP and ASAS-SN objects with a observed to catalogue V magnitude difference of $> 1\sigma$ were removed. Finally, all ASAS-SN stars were visually inspected, and 173 stars not displaying a clear rotational modulation were discarded. Combined, this gives a full catalogue of 1,079 stars from the SuperWASP and ASAS-SN catalogues.

%__________________________________________________________________

\subsection{Positional coincidence}

SuperWASP has a large pixel size of 13.7$^{\prime \prime}$ pixel$^{-1}$. When combined with the 2.5 pixel extraction aperture for photometry, the error radius is 34$^{\prime \prime}$. Although XMM-Newton sources have low positional uncertainties, there is still the possibility of chance positional coincidences between the photometric data and X-ray sources.

The SuperWASP all-sky survey is comprised of 2$\times$8 cameras in the Northern and Southern hemisphere, covering 7712 square degrees. We have light curves of 3,091,808 unique objects with detectable photometric periods from the \textit{SuperWASP periodicity} catalogue. Therefore, the mean linear separation between nearest neighbours is 180$^{\prime \prime}$ for SuperWASP objects. The 4XMM-DR9 catalogue contains 550,124 unique X-ray sources covering 1089 square degrees, overlapping the SuperWASP sky area. The mean error radius of XMM-Newton positions is $<$2$^{\prime \prime}$. This corresponds to a 0.6\% chance of an XMM-Newton source and a SuperWASP source coinciding at random. We find 39,367 positional matches (of which we use 581 to characterise the rotation-activity relation), rather than $\sim$2316 matches expected as a result of chance, suggesting that (39,367-2316)/39,367 = 94\% of the positional coincidences are the result of genuine X-ray emissions from SuperWASP sources. 

Using the same method, the mean linear separation between nearest neighbours for ASAS-SN sources is 103$^{\prime \prime}$, and a 2\% chance that XMM-Newton and ASAS-SN sources coincide at random. We find that 80\% of the positional coincidences are the result of genuine X-ray emissions from ASAS-SN sources.

%__________________________________________________________________

\subsection{Colours and reddening}
\label{colourreddening}

\begin{figure}
   \centering
   \includegraphics[width=\hsize]{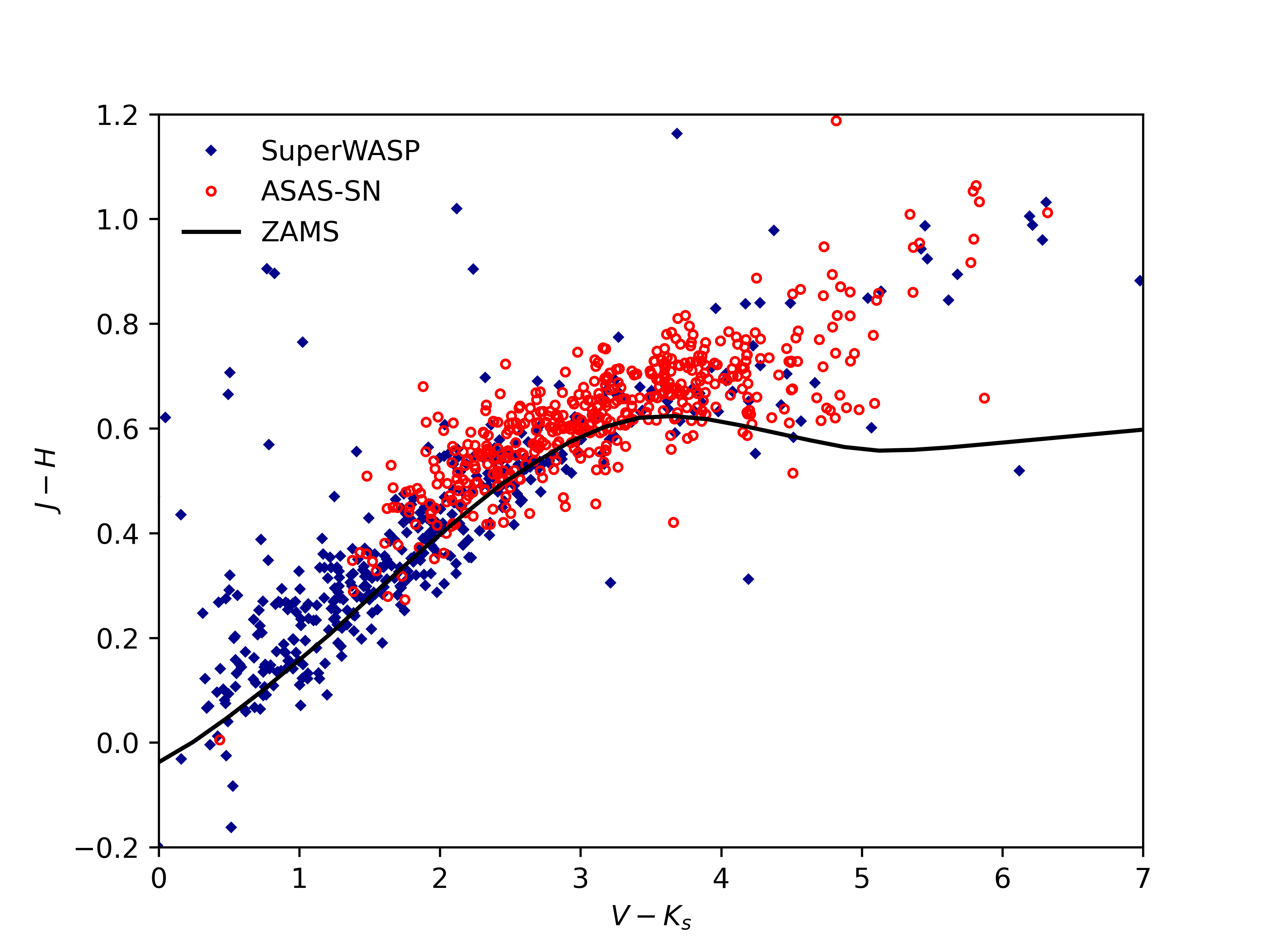}      
   \caption{The $V - K_s$, $J - H$ colour-colour plot of 900 stars in the full catalogue. The black line indicates the locus of the zero age main sequence for zero reddening.}
         \label{zams}
\end{figure}

Figure \ref{zams} shows the $V - K_s$ vs $J - H$ colour-colour plot of the catalogue, and the solid black line shows the locus of the Main Sequence for zero reddening. Extinction will move the line upwards in $J - H$ and right in $V - K_s$ according to \citet{Wegner}. There is a very slight offset shifting $J - H$ upwards, which may be due to the non-standard V magnitude calculated by SuperWASP \citep{norton2007}, however since the majority of the objects are close to the zero-reddening MS, all objects are relatively close-by.

%__________________________________________________________________

\subsection{Bolometric Flux Ratio}

To calculate bolometric luminosities, all targets were converted from observational magnitudes to absolute magnitudes, and extinction corrected.

\begin{equation}
      M_v = m - 5\log \left ( \frac{d}{10} \right) + A
      \label{extinction}
\end{equation}

where $A = 3.2[E(B-V)]$ and distance is measured in parsecs.

The extinction for each star was synthesised using the Binary and Stellar Evolution and Population Synthesis (BiSEPS) implementation of extinction given by \citet{drimmel}, using \citet{bailer2018} Gaia-DR2 distances.

Bolometric corrections were calculated from $V - K_s$ colour and \citet{bailer2018} Gaia-DR2 distances, allowing for the calculation of bolometric luminosities using $M_{bol} = BC_v - M_v$. The bolometric luminosity was calculated using

\begin{equation}
      L_{bol}=L_0 \times 10^{-0.4M_{bol}}
      \label{bolometric}
\end{equation}

where $L_0$ is the zero point luminosity 3.0128$\times10^{28}$ W. X-ray fluxes were converted to X-ray luminosities using soft X-rays and Gaia-DR2 \citet{bailer2018} distances using $L_x =4\pi D^{2}F_x$. Finally, the X-ray-to-bolometric luminosity ratios, $R_x=L_x/L_{bol}$, were determined.

\subsection{Rossby number}

We calculated the convective turnover time, $\tau$, using the $V - K_s$ colour relation derived by \cite{wright2018}. Previous empirically derived relationships have been found, e.g. \citet{Pizzolato2003} and \citet{wright2011}, however these relationships have small $V-K_s$ ranges. \citet{wright2018} extended this relationship, and as such, it is well constrained for $1.1 < V - K_s < 7.0$. It is worth noting that to derive this empirical form of the relation, \cite{wright2018} assumes a rotation-activity relation as the input, hence it might be reasonable to assume that the output will be self-consistent and will show a rotation-activity relation in line with canonical values. This relationship has the form:

\begin{equation}
      \log\tau = 0.64(\pm_{0.12}^{0.10}) + 0.25(\pm_{0.07}^{0.08}) [V - K_s]
      \label{logtau}
\end{equation}

where $\tau$ is the convective turnover time. Rossby numbers, $Ro = P_{rot}/\tau$, were calculated for all targets. 

%__________________________________________________________________

\begin{figure*}
   \centering
   \includegraphics[width=\hsize]{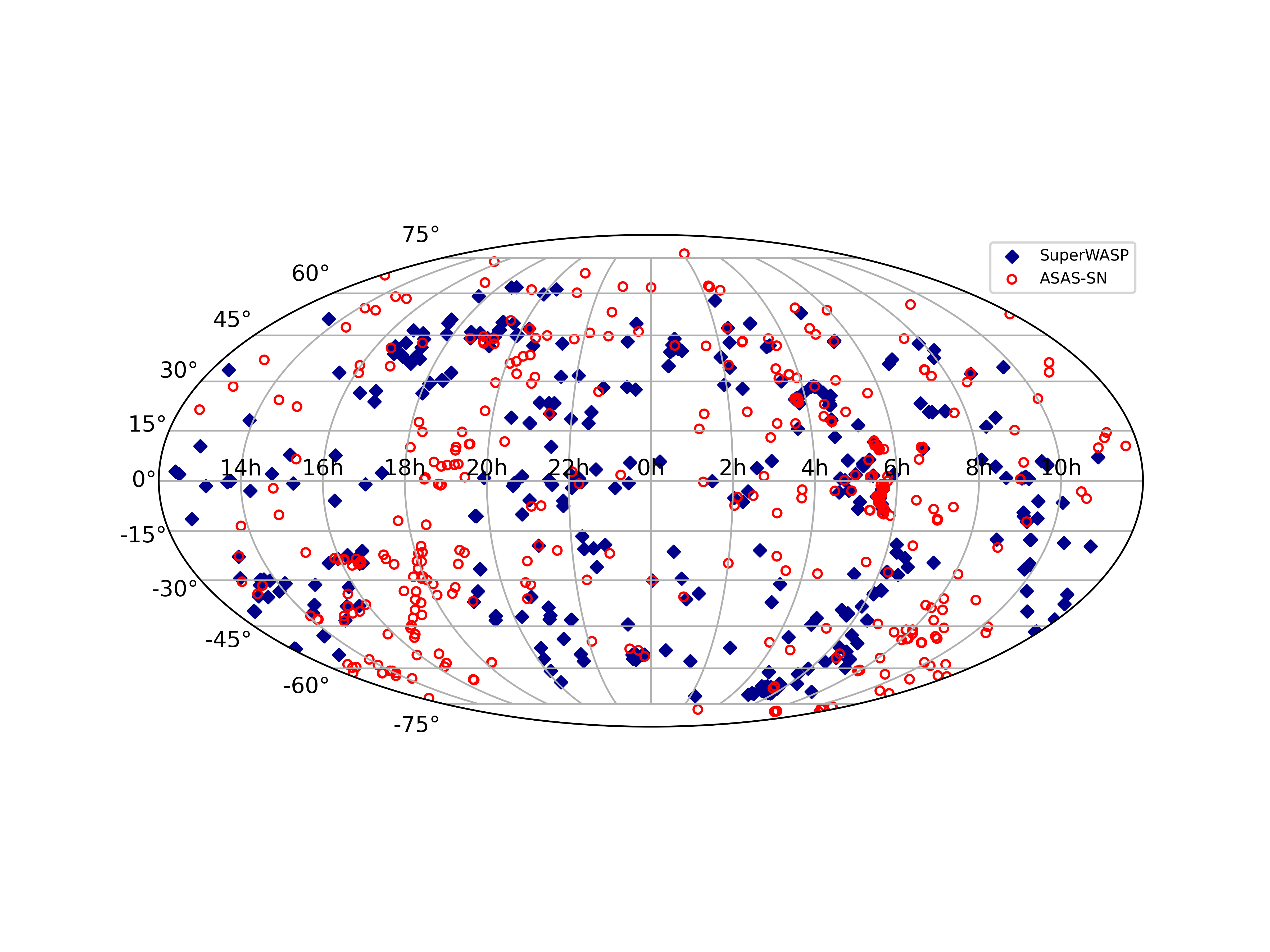}
      \caption{The spatial distribution of the 900 objects in the full catalogue coincident with 4XMM-Newton sources.}
         \label{ra_dec}
\end{figure*}

\subsection{The full catalogue}

Having characterised $Ro$ and $L_x/L_{bol}$, cuts were made to the catalogue of 1,079 stars to make it suitable for use for the rotation-activity relation. Any stars with a period of $\geq$50 days were removed, since long period photometric variability is difficult to correctly define for long periods. Late-type stars with a bolometric luminosity of $L_{bol}\geq10L_{\odot}$ were discarded as giants. The 4XMM-DR9 variability flag, \textit{$VAR\_FLAG$} was used to remove the 34 GKM-type sources detected as variable.

\begin{figure}
   \centering
   \includegraphics[width=\hsize]{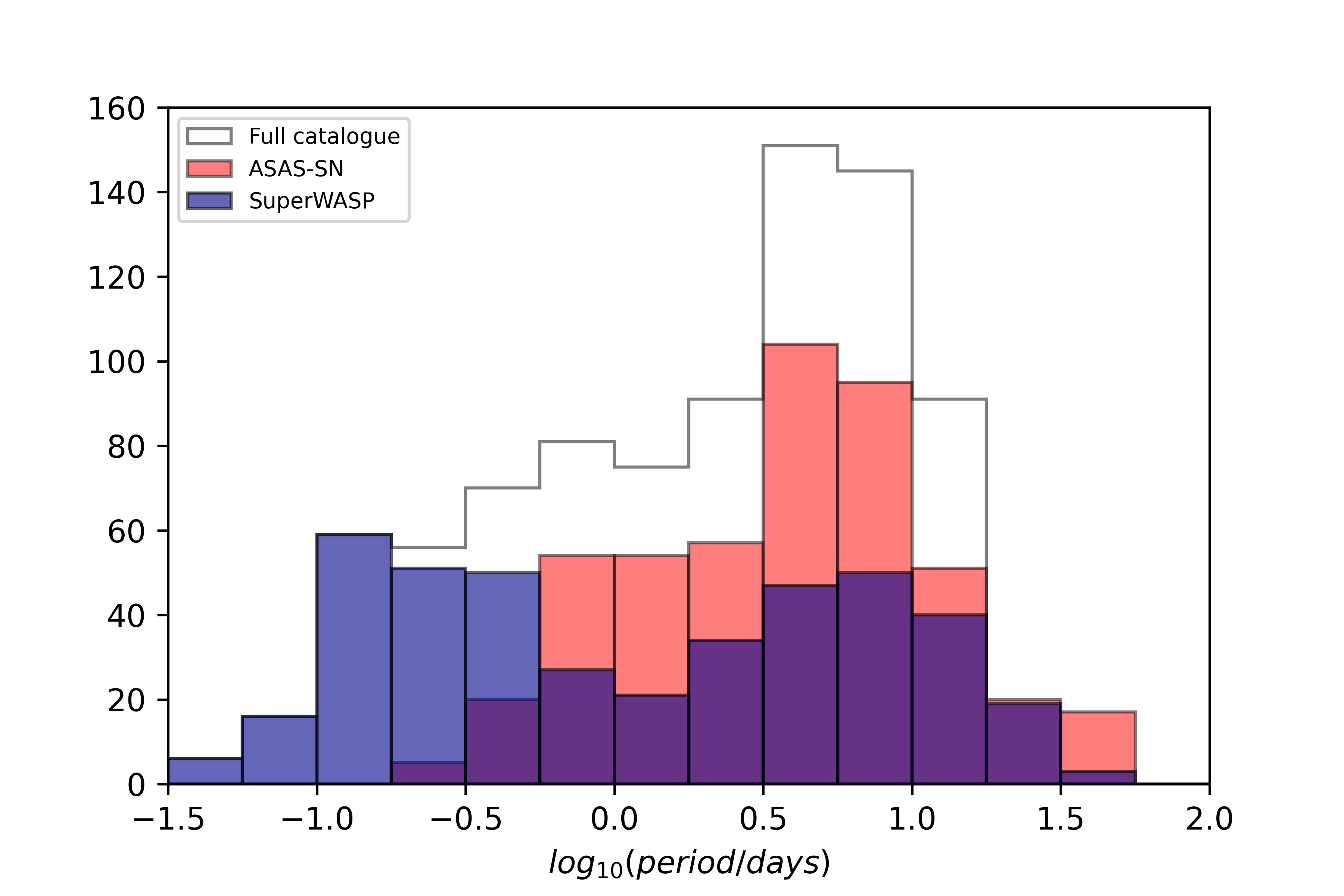}
      \caption{The period distribution of 900 stars in the full catalogue coincident with 4XMM-Newton sources.}
         \label{period}
\end{figure}

\begin{figure}
   \centering
   \includegraphics[width=\hsize]{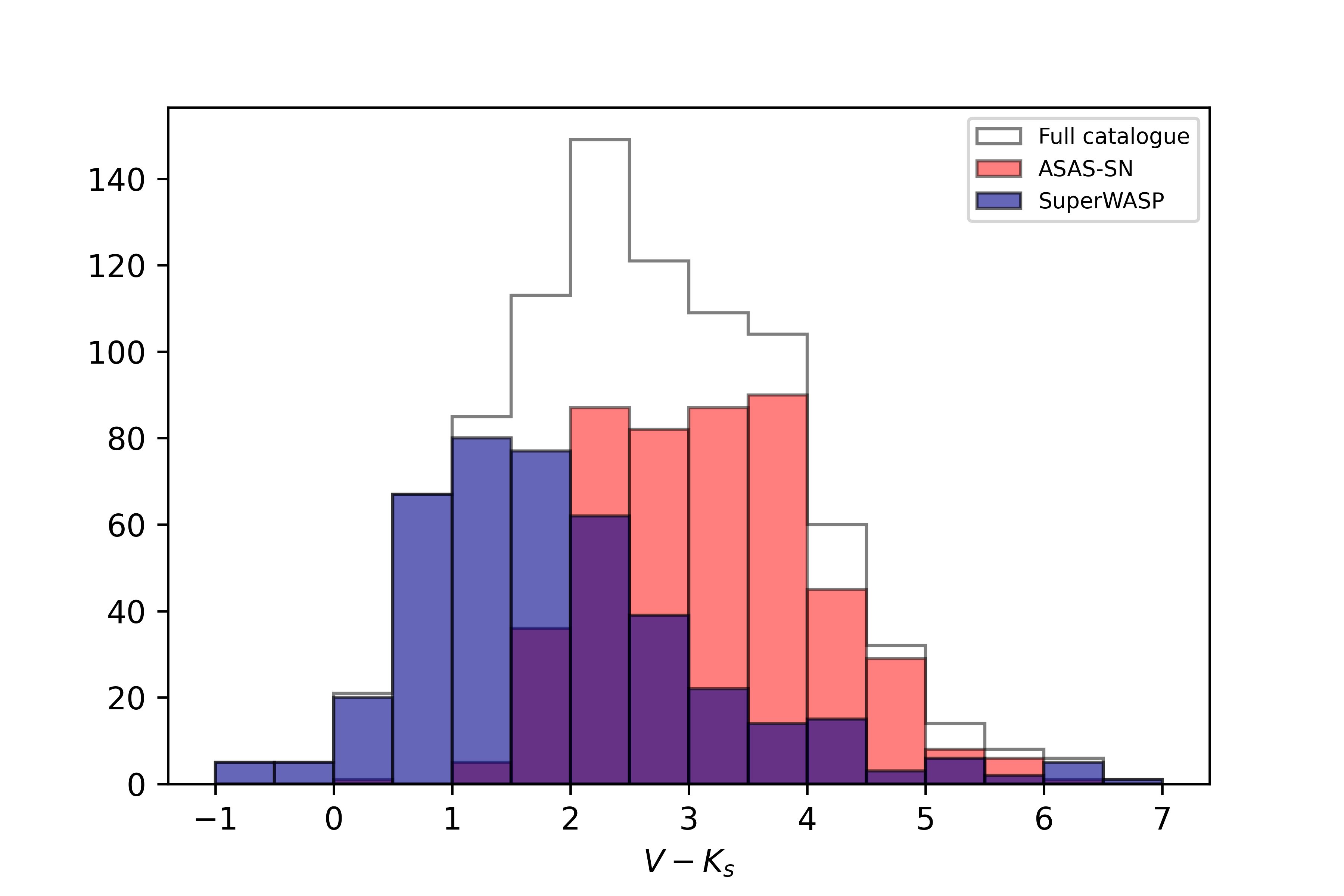}
      \caption{The $V - K_s$ colour distribution of 900 stars in the full catalogue coincident with 4XMM-Newton sources.}
         \label{colour}   
\end{figure}

The full catalogue contains 900 X-ray visible unique objects displaying rotational modulation in their photometric variability. Figure \ref{period} and Figure \ref{colour} show the distribution of period and $V-K_s$ colour in the full catalogue. The catalogue is concentrated around objects with a rotation period of 3-10 days. Like the ASAS-SN Catalogue of Variable Stars, the \textit{SuperWASP periodicity catalogue} reports the rotation period derived for each target. The periods given in this catalogue is therefore the "best guess" rotation period, and may not be the actual rotation periods of the stars. However, given that both the ASAS-SN Catalogue of Variable Stars and the \textit{SuperWASP periodicity catalogue} have independently derived almost identical rotation periods through different methods, we can be confident enough in their accuracy to use this new catalogue to probe the rotation-activity relation. Table \ref{table_wasp} and Table \ref{table_asas} show a sample of the full catalogue. Splitting the catalogue into spectral types based on $V-K_s$ colour and tables by \citet{Mamajek}, gives 431 O-A-type, 33 F-type, and 402 GKM-type stars.

The spatial distribution of the full catalogue is shown in Figure \ref{ra_dec}, indicating SuperWASP stars lying away from the Galactic Plane, and ASAS-SN stars lying mostly along the Galactic Plane.

The columns of the tables are as follows: 1. IAUNAME: The XMM-Newton identifier in the form 4XMM Jhhmmss.ss+ddmmss.s; with the position encoded in this identifier. 2. WASP ID: The SuperWASP identifier in the form 1SWASP Jhhmmss.ss+ddmmss.s; the position encoded in this identifier is identical to the position of the corresponding object in the USNO B1 catalogue, or ASAS-SN ID: The ASAS-SN identifying in the form ASASSN-V Jhhmmss.ss+ddmmss.s. 3. Spectral Type: Spectral type taken from \citet{Mamajek} based on $V - K_s$ colour. 4. V magnitude: V is the mean SuperWASP magnitude defined as $-2.5 log_{10}(\frac{F}{10^6})$ where F is the mean SuperWASP flux in microVegas - a pseudo-V magnitude which is comparable to the Tycho V magnitude, or V magnitude: ASAS-SN V magnitude. 5. Period: The SuperWASP photometric period in days identified by the \textit{SuperWASP periodicity} period search, or Period: The ASAS-SN photometric period in days. 6. Distance: Distance in parsec taken from \citet{bailer2018}. 7. $V - K$: $V - K$ calculated using 2MASS and USNOB1 catalogue magnitudes. 7. Fx: Soft X-ray flux taken from \textit{SC\_EP\_9\_FLUX} in the broad energy band, corresponding to 0.2 - 4.5 keV. 12. $\log Ro$: Rossby number calculated using relationship derived in \citet{wright2018}). 13. $\log Rx$: Logarithm of the ratio of X-ray to bolometric luminosity.

%__________________________________________________________________

\section{CHARACTERISING THE ROTATION-ACTIVITY RELATION}

To characterise the rotation-activity relation for each spectral type, we fit a power-law function  to a plot of $\log(L_x/L_{bol})$ to $\log R_o$ for the unsaturated and supersaturated regimes, where applicable, and found a mean for the saturated regime. The saturation point, $Ro_{sat}$ is initially taken to be $Ro_{sat} = 0.13$ based on canonical values e.g. \cite{wright2011}. Points for supersaturation fit are based on literature values. The fits for each regime and spectral type are based on Equation \ref{2} and an initial fit using OLS was done to identify the saturation point as $Ro_{sat} = 0.14\pm0.03$. Each regime's fit is improved using the Python package \textit{emcee}, a Markov chain Monte Carlo (MCMC) implementation of the affine-invariant ensemble sampler created by \cite{emcee}, which takes both uncertainties in $Ro$ and $(L_x/L_{bol})$ into account. 

The commonly used two-part power-law for the unsaturated and saturated regime (e.g. \cite{Pizzolato2003}, \cite{wright2011}, \cite{wright2018}) has the form:

\begin{equation}
  \frac{L_x}{L_{bol}} =
    \begin{cases}
      CRo^\beta & \text{if $Ro > Ro_{sat}$}\\
      (\frac{L_x}{L_{bol}})_{sat} & \text{if $Ro \leq Ro_{sat}$}\\
    \end{cases}  
    \label{2}
\end{equation}

It is assumed that the supersaturated regime will have a similar, but positive, power-law form, $\beta_{s}$.

%__________________________________________________________________

\section{RESULTS}

\begin{figure}[ht]
   \centering
   \includegraphics[width=\hsize]{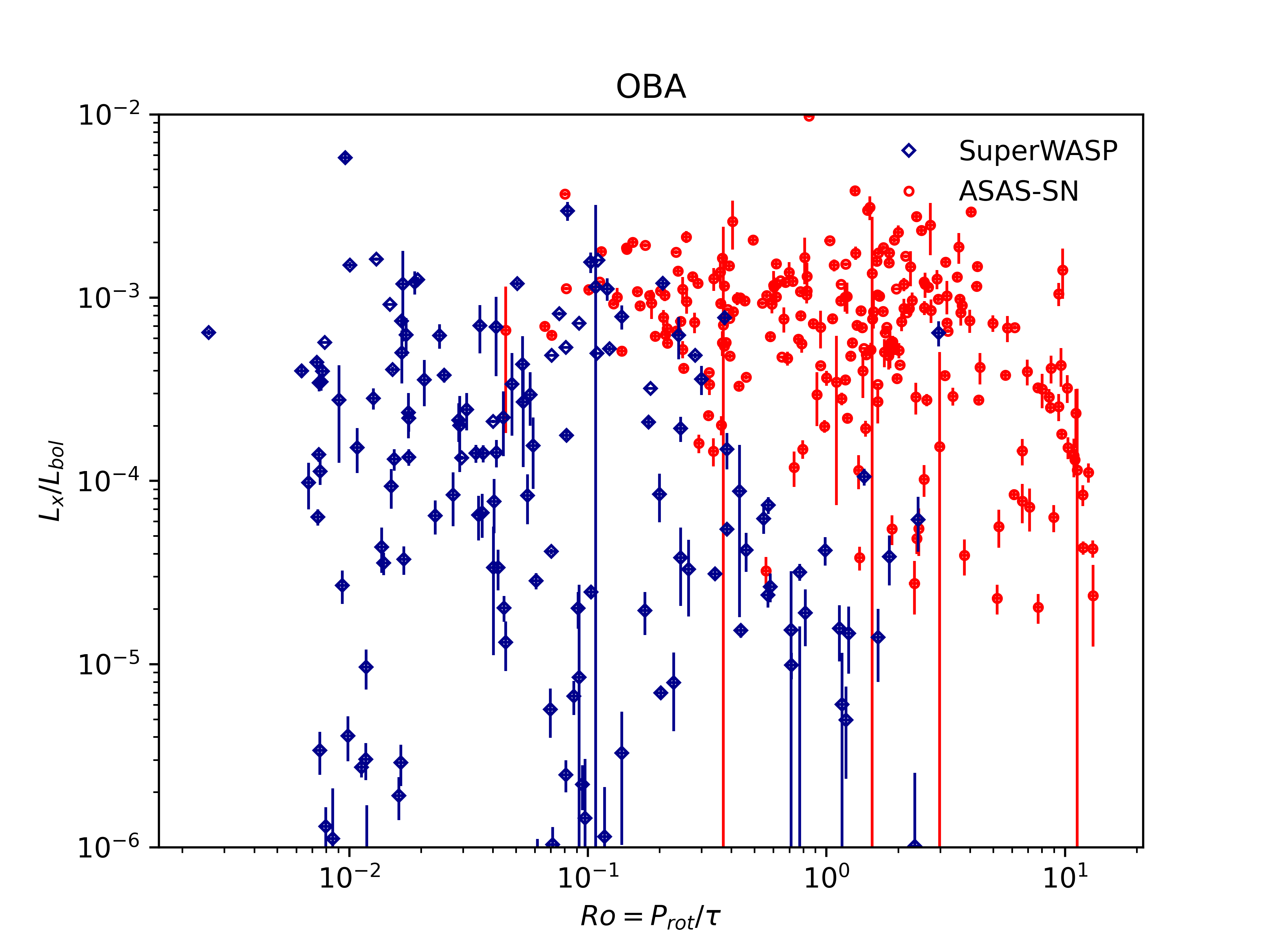}
      \caption{The 431 OBA-type stars in the catalogue do not display the rotation-activity relation.}
         \label{OBA}
\end{figure}

\begin{figure}[ht]
   \centering
   \includegraphics[width=\hsize]{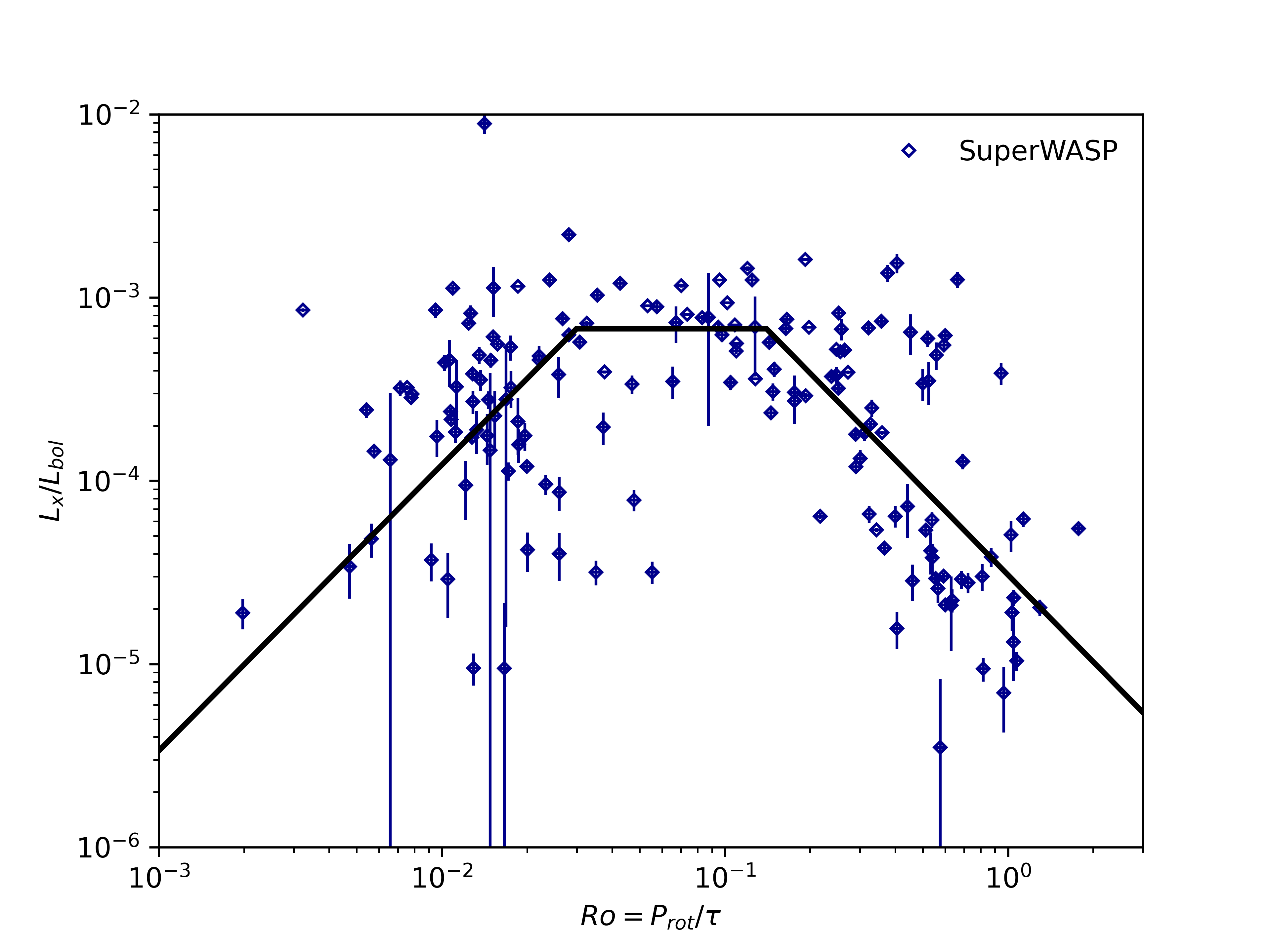}
      \caption{Fractional X-ray luminosity,$L_x/L_{bol}$, plotted against the Rossby number, $Ro$, for 192 G- to M-type SuperWASP stars. Regime thresholds are as follows: supersaturated regime ($0.03 > Ro$), saturated regime ($0.03 < Ro < 0.14 $), and unsaturated regime ($0.14 < Ro < 2.00 $)). GKM-type SuperWASP (full black line) and ASAS-SN catalogues (dashed black line).}
         \label{WASP_GKM}
\end{figure}

\begin{figure}[ht]
   \centering
   \includegraphics[width=\hsize]{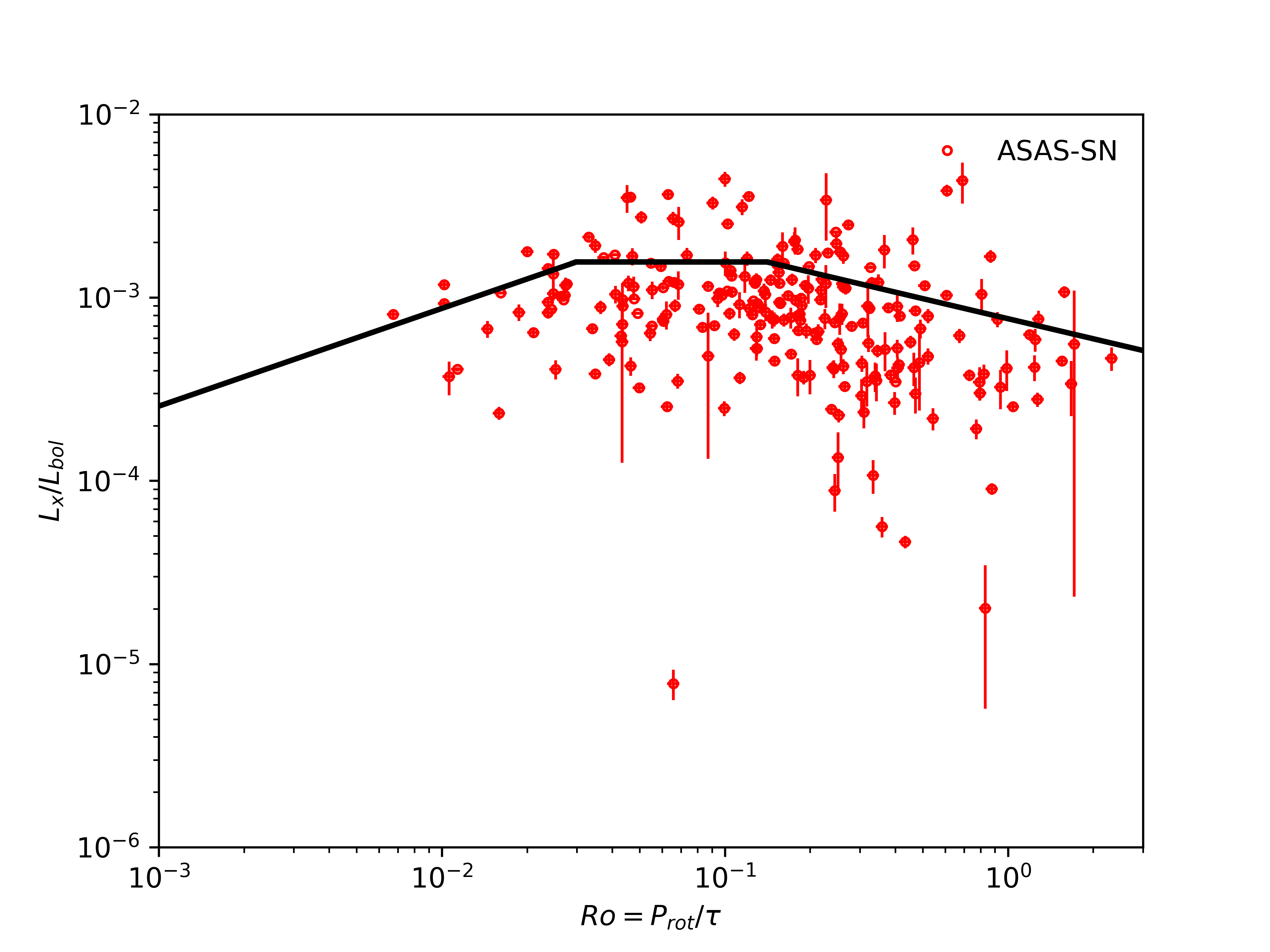}
      \caption{Fractional X-ray luminosity,$L_x/L_{bol}$, plotted against the Rossby number, $Ro$, for 244 G- to M-type ASAS-SN stars. Regime thresholds are as follows: supersaturated regime ($0.03 > Ro$), saturated regime ($0.03 < Ro < 0.14 $), and unsaturated regime ($0.14 < Ro < 2.00 $)).}
         \label{ASAS_GKM}
\end{figure}

Plots of the Rossby number, $Ro = P_{rot}/\tau$, against bolometric flux ratio, $L_x/L_{bol}$, for for the full SuperWASP and ASAS-SN catalogue are shown in Figures \ref{OBA} and \ref{WASP_GKM} and \ref{ASAS_GKM}. O-A-type stars are known to not display the rotation-activity relation, as shown in Figure \ref{OBA}. A-type stars are typically X-ray dark, whereas O- and B-type stars generate X-rays via wind shocks or for binaries, shocks between colliding winds. Any X-ray emission from A-type stars is most likely the emission from an unresolved companion star \citep{2007A&A...475..677S}.

The full catalogue was separated into the component ASAS-SN and SuperWASP catalogues and fitted separately, shown in Figure \ref{WASP_GKM} and Figure \ref{ASAS_GKM}. This demonstrates the significant difference between each subset of stars in the full catalogue. A fit to the SuperWASP catalogue shows the classical two-part power-law fit of unsaturated and saturated regime, and a third supersaturated regime, whereas the fitting to the ASAS-SN catalogue is extremely unconvincing and is shown purely to demonstrate this fact. The breakdown of fitting parameters for each regime and catalogue subset are shown in Table \ref{table:1}.

The unsaturated regime typically lies between $0.10 < Ro < 2.00 $. We find the unsaturated regime for our catalogue to be in the range $0.14 < Ro < 2.00 $. In the unsaturated regime there is an anti-correlation between the Rossby number, $Ro$, and X-ray luminosity ratio, $L_x/L_{bol}$.  We used 89 SuperWASP G- to M-type stars to fit find a power-law fit of $\beta = -1.84\pm0.18$, matching the canonical value $\beta = -2$, within uncertainties.

The saturated regime exists for fast rotating solar and late-type stars with $0.03 < Ro < 0.14$, independent of spectral type. The SuperWASP subset contains 31 saturated stars. Previous studies have found the value of the saturation point to be between $Ro_{sat}\sim0.08$ \citep{marsden2009} and $Ro_{sat}\sim0.16\pm0.02$ \citep{wright2011}. We find the value of $Ro_{sat}=0.14\pm0.03$, and find a mean saturation of $\log(L_x/L_{bol}) = -3.14$.

The supersaturated regime occurs for the very fastest rotating stars with $Ro < 0.03$. For the fastest rotating stars, $L_x/L_{bol}$ has been observed to decrease below the saturation level. Our SuperWASP G- to M-type sample clearly displays evidence of a supersaturated regime, with a power-law slope fit of $\beta_{s} =  1.42\pm0.26$, with 60 stars in this subset. 

%__________________________________________________________________
% OBA
%__________________________________________________________________

%__________________________________________________________________
% TABLE BOTH
%__________________________________________________________________

\begin{table*}
\caption{Parameters for the GKM-type regime, separated by catalogues and period ranges. nWasp and nASAS-SN indicate the number of stars in each regime.}
\label{table:1}
\centering 
\begin{tabular}{c c c c c c} 
\hline\hline 
  Regime & nWASP & WASP & nASAS-SN & ASAS-SN \\ 

\hline                        
 Supersaturated ($\beta$) & 60 & 1.42$\pm$0.22 &  21 & 0.43$\pm$0.24   \\
 Saturated (Rx)  & 31 & -3.14 &   76 & -2.81  \\ 
 Unsaturated ($\beta_{s}$)  & 89 & -1.84$\pm$0.18 &  125 & -0.33$\pm$0.11  \\ 
\hline    
\end{tabular}
\end{table*}

The scatter in $L_x/L_{bol}$ in the saturated regime is likely to arise from a number of factors: X-ray luminosity varying over a stellar cycle (sometimes up to an order of magnitude in our Sun \citep{peres2000}), and with a significant portion of X-ray sources having only one detection, varying starspot coverage may change the measured activity of the star. 

In $V-K_s$ space, the stars used to characterise the rotation-activity relation are distributed across the range $1.1 \leq V-K_s \leq 5.7$, with K-type stars dominating. Uncertainties in V magnitudes are generally small ($\sim3$\%) compared to the measured X-ray fluxes, which have a typical uncertainty of $\sim20$\%. Distance uncertainties are also typically small ($\sim5$\%). As such, in $L_x/L_{bol}$, the X-ray uncertainty dominates. The rotation periods for the SuperWASP archive, $P_{rot}$ are derived empirically, and the typical uncertainty is $\sim0.1$\%. It is assumed that ASAS-SN periods have an equivalent uncertainty. Additionally, the SuperWASP subset has marginally smaller uncertainties, with a mean uncertainty of $\pm6.05\times10^{-5}$ in ${L_x}/{L_{bol}}$ compared to $\pm9.88\times10^{-5}$ for the ASAS-SN subset. Similarly, in $R_o$, the SuperWASP catalogue has a mean uncertainty of $\pm$0.06, whereas the ASAS-SN catalogue has a mean uncertainty of $\pm$0.15.

%__________________________________________________________________

\section{DISCUSSION}

Somewhat unexpectedly, the ASAS-SN and SuperWASP catalogues show very different results. Notably, the SuperWASP catalogue shows convincing evidence for the unsaturated and supersaturated regimes, and with a higher saturation point, whereas the ASAS-SN catalogue shows no convincing evidence for the unsaturated or supersaturated regimes.

It is likely even after cleaning the data, both subsets still have some level of contamination. \citet{2018asasJ} states that rotational variables are "usually the most difficult to precisely classify", and as such rotational variables in the ASAS-SN Catalogue of Variable Stars and the SuperWASP periodicity catalogue may be unclassified binaries or low amplitude Cepheids, especially for fast rotating objects with periods of $\leq$ 1 day, or for longer period objects. \citet{2018asasJ} notes that the majority of rotational variables in the catalogue overlap with pulsators in colour and period space. Without spectroscopic measurements for this study, these light curves can still be open to interpretation. However, since both catalogues have been classified by eye using the same method (for initial selection of the SuperWASP objects, and for final verification of the ASAS-SN objects), contamination alone cannot account for the inter-catalogue variation.

\begin{figure}[ht]
   \centering
   \includegraphics[width=\hsize]{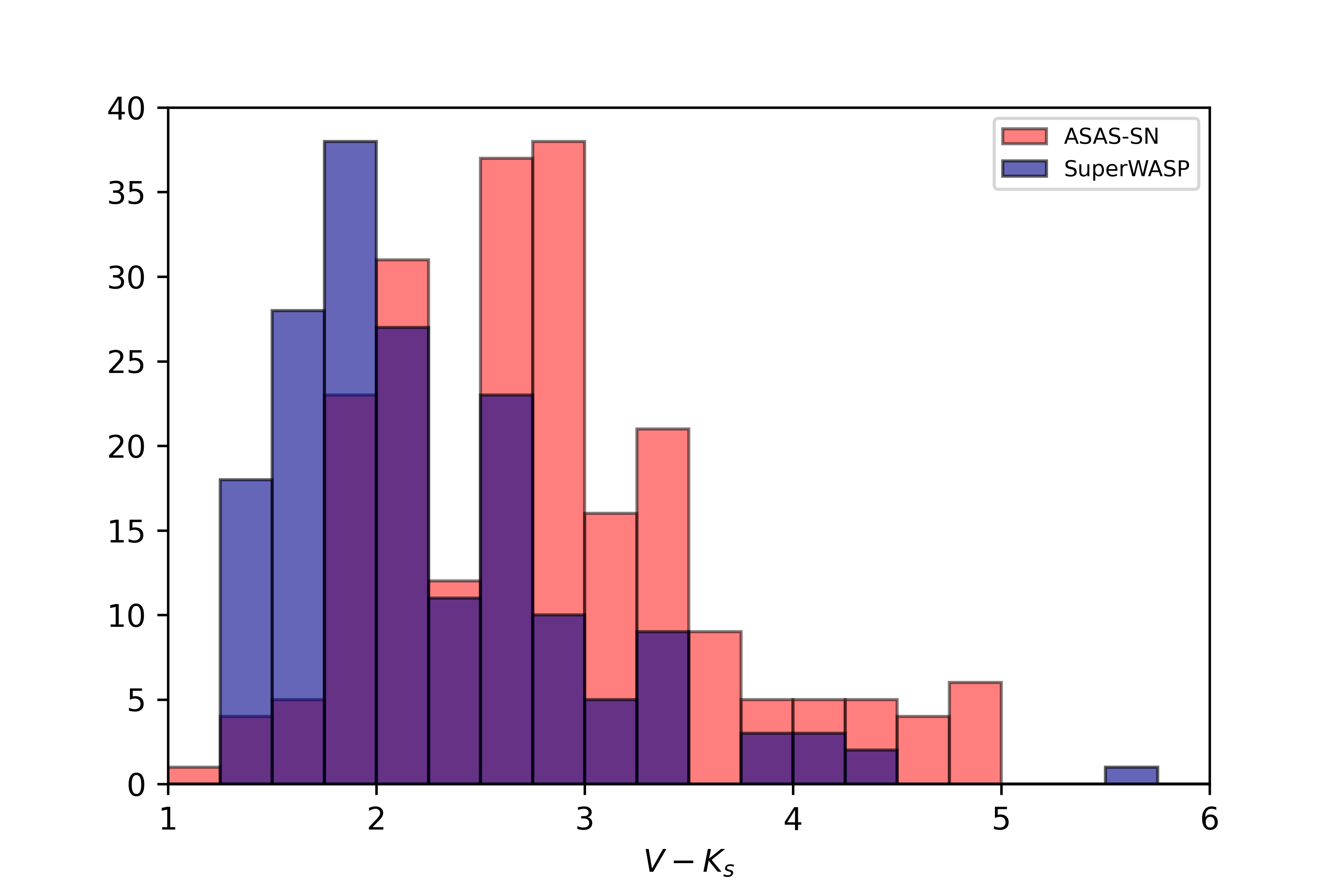}
      \caption{The $V-K_s$ colour distribution of GKM-type stars.}
         \label{hist_v-k_GKM}
\end{figure}

Investigating the differences between the two catalogues shows that the data sets show similar distributions of X-ray flux, however the ASAS-SN catalogue shows a distribution towards redder, fainter objects, shown in Figure \ref{hist_v-k_GKM}. This is unsurprising, since the limiting magnitude of ASAS-SN is V = 17 mag, whereas SuperWASP is V = 15 mag, however we would still expect such objects to follow the relationship.

\begin{figure}[ht]
   \centering
   \includegraphics[width=\hsize]{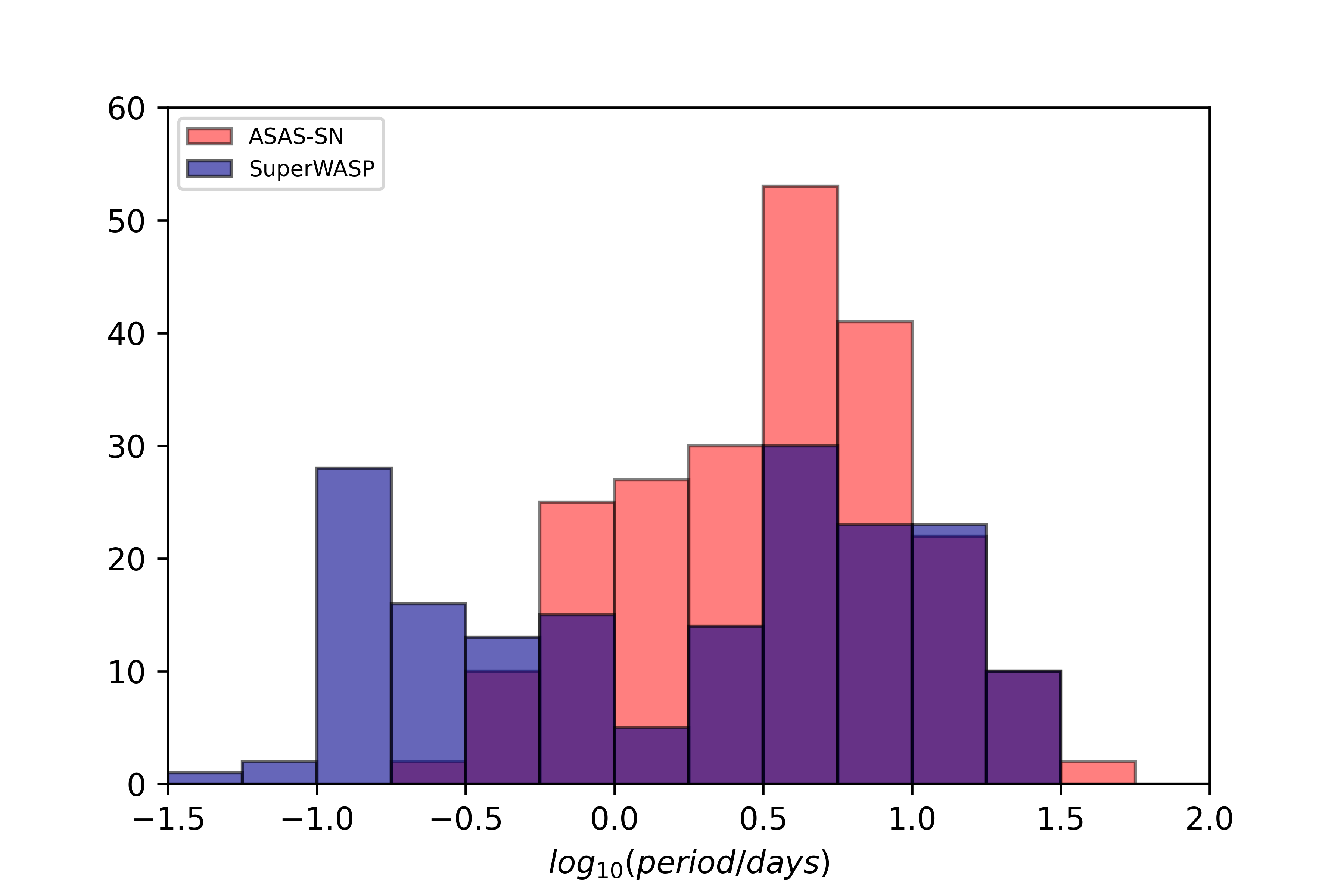}
      \caption{The period distribution of GKM-type stars.}
         \label{hist_period_GKM}
\end{figure}

Figure \ref{hist_period_GKM} shows that the ASAS-SN catalogue contains no GKM-type objects with a period $\leq$0.3 days, whereas almost 25\% of the SuperWASP catalogue contains objects within that period range. All flagged SuperWASP objects were removed from the catalogue to avoid the inclusion of spurious periods, hence there is a lack of SuperWASP objects with a period of close to 1 day. The SuperWASP data set includes a larger proportion of short period objects, made very clear by Figure \ref{OBA} which shows a significant difference between Ro, and hence period range, of the two catalogues. As a test, removing all ASAS-SN objects with a period of $\geq$ 10 days does improve the power-law fit slightly, to $\beta=-0.87\pm0.18$. It is likely that the detection and automated classification methods used by ASAS-SN struggles with the period detection and classification of faint, long period rotational objects. We can find no other conclusive reason why the ASAS-SN sample appears to not follow the rotation-activity relation.

It is likely that misclassification of rotationally-modulated objects is also the primary reason for the small difference for the unsaturated regime power-law fit between the canonical value of $\beta=-2$ and for the SuperWASP value, which finds $\beta=-1.84\pm0.18$. Another, less likely possibility is that field stars may follow a marginally different rotation-activity relation than stars in clusters. Stars in clusters will have a similar age, metallicity, composition, and formation, and would experience magnetic braking and activity decline at similar rates, whereas field stars may not, however this is beyond the scope of this study.

%__________________________________________________________________

\section{CONCLUSION}

The study of the rotation-activity relation has so far been limited by the availability of deep observations over long baselines, and accuracy of stellar rotation periods. We sought to tackle this through the use of the SuperWASP archive, which contains $\sim$10 years of high-cadence, long baseline observations, and a period search of the resulting $>$ 30 million light curves, and the automatically classified light curves in the ASAS-SN Catalogue of Variable Stars.

Through cross-correlations of SuperWASP and ASAS-SN photometric data and XMM-Newton X-ray data, we have updated the work by \citet{norton2007}, and we have characterised 900 rotationally modulated stars with photometric rotation periods from the \textit{SuperWASP periodicity} period search and ASAS-SN Catalogue of Variable Stars, of which 402 late-type main-sequence field stars are used to study the rotation-activity relation.

We find that while the ASAS-SN Catalogue of Variable stars provides a large catalogue of stars classified as rotationally modulated, using the catalogue without visually inspecting each light curve means that automatically classified data cannot be used to characterise the rotation-activity relation, and even after visual inspection, there may still be issues with the period detection for long period rotational objects in ASAS-SN. Having classified the SuperWASP light curves by eye, we find that late-type stars follow the rotation-activity relation, and the slope of the power-law fit of $\beta = -1.84\pm0.18$ is in line with the canonical value of $\beta = -2$. We also find a third supersaturation regime for the fastest rotating stars, and fit a power-law slope of $\beta_{s} = 1.42\pm0.26$. We suggest that the inclusion of indistinguishable non-rotational objects is the main reason for the lower $\beta$ fit.

Since the rotation-activity relation is believed to be a proxy for the stellar magnetic dynamo, and most dynamo models cannot yet reliably predict a correlation between rotation and activity indicators such as X-ray activity, all-sky observational studies prove vital for understanding the stellar dynamo, however we advise that automatically classified rotationally modulated light curves are first thoroughly vetted by eye prior to avoid contamination of the sample.

%__________________________________________________________________

\begin{acknowledgements}

We would like to thank the anonymous referee for their time and their constructive comments which helped to improve the quality of the paper.

This work was supported by the Science and Technology Facilities Council [grant number ST/P006760/1] through the DISCnet Centre for Doctoral Training.

This research has made use of the TOPCAT and STILTS software packages (written by Mark Taylor, University of Bristol).

WASP-North is hosted by the Isaac Newton Group on La Palma and WASP-South is hosted by the South African Astronomical Observatory (SAAO) and we are grateful for their ongoing support and assistance. Funding for WASP comes from consortium universities and from the UK’s Science and Technology Facilities Council.

This research has made use of data obtained from the 4XMM XMM-Newton serendipitous source catalogue compiled by the 10 institutes of the XMM-Newton Survey Science Centre selected by ESA.

\end{acknowledgements}

%__________________________________________________________________

\bibliographystyle{pasa-mnras}
\bibliography{bibliography}

%__________________________________________________________________

\begin{sidewaystable*}
\caption{Subset of the 180 G- to M-type SuperWASP objects from the full catalogue containing 900 X-ray visible unique objects displaying rotational modulation in their photometric variability.}
    \scalebox{0.9}{
\begin{tabular}{llllllllll}
\hline\hline

XMM ID   & WASP ID   & Spectral Type & Period & V & V-K & Distance   & F$_x$ & $\log Ro$ & $\log R_x$  \\
  &    & & [d] &  [mag] &  & [pc] & [$10^{-14} erg/cm^2/s$]  & &  \\
\hline
4XMM J000234.8-300525  &  1SWASPJ000234.82-300525.3  &  K4.5V  & 0.71 & 13.81 & 2.78 & 23.25 & 5.46E+29 $\pm$ 1.40E+28 &  -1.49 $\pm$ 0.07  &  -4.63 $\pm$ 0.15 \\ 
4XMM J001321.2+054505  &  1SWASPJ001321.26+054505.8  &  G8V    & 13.39 & 11.63 & 1.73 & 62.56 & 1.02E+30 $\pm$ 6.81E+28 &  0.05  $\pm$ 0.00   &  -5.24 $\pm$ 0.49 \\ 
4XMM J002847.8+345355  &  1SWASPJ002841.59+345358.9  &  G2V    & 27.94 & 14.73 & 1.47 & 104.25 & 4.81E+29 $\pm$ 1.56E+29 &  0.44  $\pm$ 0.02   &  -4.22 $\pm$ 1.42 \\ 
4XMM J002914.0+345133  &  1SWASPJ002918.85+345102.9  &  K1V    & 27.83 & 12.83 & 2.06 & 41.83 & 3.42E+29 $\pm$ 3.99E+28 &  0.29  $\pm$ 0.01   &  -4.83 $\pm$ 0.62 \\ 
4XMM J003823.9+401250  &  1SWASPJ003823.99+401250.2  &  K2V    & 4.66 & 12.40 & 2.15 & 92.82 & 2.37E+31 $\pm$ 2.57E+29 &  -0.51 $\pm$ 0.03  &  -4.05 $\pm$ 0.44 \\ 
4XMM J005305.0+394730  &  1SWASPJ005310.20+394834.0  &  G7V    & 25.88 & 13.00 & 1.68 & 51.32 & 2.61E+29 $\pm$ 5.99E+28 &  0.35  $\pm$ 0.02   &  -4.87 $\pm$ 1.13 \\ 
4XMM J011858.9-341427  &  1SWASPJ011858.94-341428.5  &  K1V    & 15.78 & 12.98 & 2.06 & 68.63 & 7.94E+30 $\pm$ 1.18E+29 &  0.04  $\pm$ 0.00   &  -4.44 $\pm$ 0.21 \\ 
4XMM J015636.6+285643  &  1SWASPJ015636.65+285643.7  &  K3.5V  & 5.27 & 12.94 & 2.63 & 22.16 & 6.72E+29 $\pm$ 3.02E+28 &  -0.58 $\pm$ 0.03  &  -4.59 $\pm$ 0.23 \\ 
4XMM J020930.9+342421  &  1SWASPJ020930.99+342421.0  &  K2V    & 3.85 & 13.08 & 2.14 & 129.04 & 2.06E+31 $\pm$ 9.67E+29 &  -0.59 $\pm$ 0.03  &  -4.04 $\pm$ 0.55 \\ 
4XMM J020953.8+342309  &  1SWASPJ020957.38+342412.6  &  K4V    & 0.40 & 13.91 & 2.69 & 19.93 & 8.60E+28 $\pm$ 1.51E+28 &  -1.71 $\pm$ 0.09  &  -4.51 $\pm$ 0.79 \\ 
4XMM J021442.0-061635  &  1SWASPJ021442.07-061635.7  &  K3V    & 18.97 & 12.48 & 2.44 & 56.72 & 7.57E+30 $\pm$ 2.31E+29 &  0.03  $\pm$ 0.00   &  -4.40 $\pm$ 0.24 \\ 
4XMM J022216.2-031036  &  1SWASPJ022215.35-031031.8  &  K3.5V  & 0.20 & 13.89 & 2.58 & 20.74 & 1.44E+28 $\pm$ 5.58E+27 &  -1.98 $\pm$ 0.10  &  -4.95 $\pm$ 1.92 \\ 
4XMM J022319.5+472720  &  1SWASPJ022319.50+472722.1  &  G4V    & 3.16 & 11.45 & 1.53 & 18.33 & 3.63E+29 $\pm$ 3.94E+28 &  -0.52 $\pm$ 0.03  &  -4.84 $\pm$ 0.53 \\ 
4XMM J022413.5+274136  &  1SWASPJ022413.54+274136.1  &  K2.5V  & 0.16 & 11.72 & 2.28 & 32.26 & 1.24E+30 $\pm$ 1.59E+29 &  -2.02 $\pm$ 0.10  &  -4.40 $\pm$ 1.00 \\ 
4XMM J023507.6+034357  &  1SWASPJ023507.59+034356.7  &  K1.5V  & 4.23 & 12.38 & 2.10 & 7.88 & 2.95E+28 $\pm$ 2.16E+27 &  -0.54 $\pm$ 0.03  &  -5.06 $\pm$ 0.37 \\ 
4XMM J023732.1-522331  &  1SWASPJ023732.16-522332.5  &  K5V    & 0.13 & 14.51 & 2.90 & 46.08 & 4.32E+29 $\pm$ 4.08E+28 &  -2.27 $\pm$ 0.11  &  -4.63 $\pm$ 0.45 \\ 
4XMM J024614.4-205011  &  1SWASPJ024622.47-205004.5  &  G8V    & 0.18 & 14.70 & 1.81 & 177.96 & 1.05E+32 $\pm$ 6.18E+30 &  -1.85 $\pm$ 0.09  &  -2.97 $\pm$ 0.36 \\ 
4XMM J025700.4+055415  &  1SWASPJ025702.21+055338.2  &  G0V    & 0.35 & 13.56 & 1.34 & 141.66 & 2.22E+30 $\pm$ 4.03E+29 &  -1.43 $\pm$ 0.07  &  -4.41 $\pm$ 0.88 \\ 
4XMM J030816.4+490050  &  1SWASPJ030816.46+490051.0  &  K4V    & 1.73 & 12.50 & 2.72 & 24.83 & 2.26E+30 $\pm$ 7.57E+28 &  -1.08 $\pm$ 0.05  &  -4.54 $\pm$ 0.17 \\ 
4XMM J030846.5+485946  &  1SWASPJ030846.52+485946.5  &  K5V    & 0.14 & 12.67 & 2.94 & 25.76 & 4.30E+29 $\pm$ 3.25E+28 &  -2.24 $\pm$ 0.11  &  -4.95 $\pm$ 0.38 \\ 
4XMM J032240.3-365917  &  1SWASPJ032246.95-370025.9  &  G3V    & 0.36 & 11.84 & 1.48 & 43.99 & 3.48E+29 $\pm$ 5.32E+28 &  -1.46 $\pm$ 0.07  &  -5.31 $\pm$ 0.82 \\ 
4XMM J032807.4-311826  &  1SWASPJ032807.62-311826.7  &  K3.5V  & 21.14 & 10.78 & 2.61 & 5.27 & 4.77E+28 $\pm$ 6.14E+27 &  0.03  $\pm$ 0.00   &  -4.99 $\pm$ 0.64 \\ 
4XMM J032814.9+300119  &  1SWASPJ032814.96+300119.1  &  G8V    & 0.16 & 11.61 & 1.78 & 22.64 & 6.59E+29 $\pm$ 3.65E+28 &  -1.89 $\pm$ 0.09  &  -4.95 $\pm$ 0.32 \\ 
4XMM J033939.5+152954  &  1SWASPJ033939.49+152954.9  &  K2V    & 4.37 & 11.82 & 2.16 & 15.54 & 2.98E+29 $\pm$ 2.28E+28 &  -0.54 $\pm$ 0.03  &  -4.83 $\pm$ 0.40 \\ 
4XMM J034322.4+242910  &  1SWASPJ034322.39+242910.6  &  K1V    & 4.49 & 12.58 & 2.08 & 23.6 & 3.43E+29 $\pm$ 3.16E+28 &  -0.51 $\pm$ 0.03  &  -4.76 $\pm$ 0.45 \\ 
4XMM J034403.5+243015  &  1SWASPJ034403.55+243015.2  &  K0.5V  & 1.48 & 10.93 & 1.98 & 13.5 & 1.99E+30 $\pm$ 1.58E+28 &  -0.97 $\pm$ 0.05  &  -4.93 $\pm$ 0.08 \\ 
4XMM J034638.7+245734  &  1SWASPJ034638.77+245734.7  &  G4V    & 2.32 & 10.39 & 1.56 & 13.63 & 2.72E+29 $\pm$ 9.97E+27 &  -0.66 $\pm$ 0.03  &  -5.60 $\pm$ 0.22 \\ 
4XMM J034902.4+231607  &  1SWASPJ034902.32+231509.0  &  K2V    & 6.15 & 11.85 & 2.16 & 13.99 & 2.11E+28 $\pm$ 4.78E+27 &  -0.39 $\pm$ 0.02  &  -5.45 $\pm$ 1.24 \\ 
4XMM J042041.1-483736  &  1SWASPJ042041.16-483735.7  &  K5.5V  & 0.38 & 13.98 & 3.04 & 37.53 & 1.17E+30 $\pm$ 7.02E+28 &  -1.82 $\pm$ 0.09  &  -4.43 $\pm$ 0.27 \\ 
4XMM J043356.3-612916  &  1SWASPJ043356.50-612916.5  &  K0.5V  & 3.39 & 10.57 & 1.98 & 9.29 & 9.51E+29 $\pm$ 1.29E+28 &  -0.61 $\pm$ 0.03  &  -5.14 $\pm$ 0.07 \\ 
4XMM J043454.5+240512  &  1SWASPJ043456.17+240543.5  &  M2V    & 0.65 & 13.68 & 4.24 & 22.73 & 2.16E+28 $\pm$ 4.25E+27 &  -1.89 $\pm$ 0.09  &  -5.73 $\pm$ 1.13 \\ 
4XMM J043558.9+223835  &  1SWASPJ043558.93+223835.1  &  M2V    & 1.93 & 12.03 & 4.28 & 15.82 & 1.98E+30 $\pm$ 2.38E+28 &  -1.43 $\pm$ 0.07  &  -5.14 $\pm$ 0.10 \\ 
4XMM J044009.2+253532  &  1SWASPJ044009.15+253533.1  &  K6.5V  & 5.61 & 10.81 & 3.24 & 4.42 & 4.39E+29 $\pm$ 4.66E+27 &  -0.70 $\pm$ 0.04  &  -5.10 $\pm$ 0.06 \\ 
4XMM J044910.5+060244  &  1SWASPJ044910.64+060243.8  &  G9V    & 10.49 & 12.06 & 1.89 & 22.68 & 7.61E+28 $\pm$ 1.24E+28 &  -0.09 $\pm$ 0.00   &  -5.31 $\pm$ 0.87 \\ 

\hline\hline
\end{tabular}
}
\label{table_wasp} 
\end{sidewaystable*}
    
%__________________________________________________________________

\begin{sidewaystable*}
\caption{Subset of the 222 G- to M-type ASAS-SN objects from the full catalogue containing 900 X-ray visible unique objects displaying rotational modulation in their photometric variability.}
    \scalebox{0.9}{
\begin{tabular}{llllllllll}
\hline\hline
XMM ID   & ASAS-SN ID   & Spectral Type & Period & V & V-K & Distance   & F$_x$ & $\log Ro$ & $\log R_x$  \\
  &    & & [d] &  [mag] &  & [pc] & [$10^{-14} erg/cm^2/s$]  & &  \\
\hline
4XMM J000228.1-300442  &  ASASSN-V J000228.21-300443.4  &  K5V  &  5.46  &  13.02  &  2.89  &  24.66  &  6.41E-14  &  -0.62 $\pm$ 0.03  &  -3.61 $\pm$ 0.17  \\
4XMM J004154.9+413332  &  ASASSN-V J004154.90+413332.3  &  K4V  &  7.71  &  15.81  &  2.76  &  164.18  &  2.86E-14  &  -0.44 $\pm$ 0.02  &  -2.74 $\pm$ 0.57  \\
4XMM J005414.1-351713  &  ASASSN-V J005414.09-351712.6  &  K4V  &  3.40  &  13.74  &  2.68  &  20.38  &  9.53E-14  &  -0.79 $\pm$ 0.04  &  -3.12 $\pm$ 0.26  \\
4XMM J011226.8+152751  &  ASASSN-V J011226.86+152752.4  &  K5.5V  &  1.16  &  14.65  &  3.06  &  24.94  &  1.64E-13  &  -1.33 $\pm$ 0.07  &  -2.45 $\pm$ 0.12  \\
4XMM J011648.3-002104  &  ASASSN-V J011648.31-002103.5  &  K2.5V  &  25.10  &  11.63  &  2.23  &  60.99  &  2.46E-13  &  0.19 $\pm$ 0.01  &  -3.35 $\pm$ 0.22  \\
4XMM J012106.0+195953  &  ASASSN-V J012106.02+195952.8  &  K3.5V  &  1.33  &  15.38  &  2.54  &  142.70  &  2.98E-14  &  -1.16 $\pm$ 0.06  &  -2.59 $\pm$ 0.53  \\
4XMM J020656.2-044905  &  ASASSN-V J020656.27-044906.1  &  G9V  &  6.01  &  14.83  &  1.87  &  62.00  &  7.13E-15  &  -0.33 $\pm$ 0.02  &  -3.52 $\pm$ 0.78  \\
4XMM J020841.2+351243  &  ASASSN-V J020841.29+351243.6  &  K5V  &  4.17  &  14.97  &  2.92  &  27.02  &  2.84E-14  &  -0.74 $\pm$ 0.04  &  -3.18 $\pm$ 0.18  \\
4XMM J021802.8+625416  &  ASASSN-V J021803.03+625418.1  &  K4.5V  &  0.55  &  14.26  &  2.84  &  24.31  &  3.29E-14  &  -1.60 $\pm$ 0.08  &  -3.39 $\pm$ 0.41  \\
4XMM J023000.7-042536  &  ASASSN-V J023000.79-042536.7  &  K8V  &  1.07  &  13.68  &  3.47  &  12.66  &  4.47E-13  &  -1.48 $\pm$ 0.07  &  -2.67 $\pm$ 0.08  \\
4XMM J025805.9+125446  &  ASASSN-V J025806.02+125446.9  &  K5.5V  &  1.16  &  14.47  &  3.08  &  27.19  &  2.88E-14  &  -1.33 $\pm$ 0.07  &  -3.37 $\pm$ 0.39  \\
4XMM J030654.5-775740  &  ASASSN-V J030654.80-775740.6  &  K4.5V  &  3.18  &  15.36  &  2.78  &  141.89  &  2.31E-14  &  -0.83 $\pm$ 0.04  &  -3.12 $\pm$ 0.36  \\
4XMM J031331.7-223658  &  ASASSN-V J031331.73-223656.8  &  K4V  &  5.36  &  15.38  &  2.71  &  177.06  &  1.87E-14  &  -0.59 $\pm$ 0.03  &  -3.11 $\pm$ 0.60  \\
4XMM J033046.8+435159  &  ASASSN-V J033046.92+435157.6  &  K5V  &  5.22  &  15.08  &  2.98  &  266.72  &  4.97E-14  &  -0.64 $\pm$ 0.03  &  -2.92 $\pm$ 0.77  \\
4XMM J033202.1-270251  &  ASASSN-V J033202.12-270251.9  &  K5.5V  &  0.66  &  12.54  &  3.06  &  14.35  &  4.52E-13  &  -1.58 $\pm$ 0.08  &  -2.99 $\pm$ 0.07  \\
4XMM J033735.1+170515  &  ASASSN-V J033735.10+170516.0  &  K6.5V  &  0.46  &  13.57  &  3.32  &  15.81  &  4.81E-14  &  -1.80 $\pm$ 0.09  &  -3.63 $\pm$ 0.30  \\
4XMM J034008.5+412846  &  ASASSN-V J034008.44+412847.0  &  K3V  &  13.70  &  13.15  &  2.45  &  65.84  &  5.81E-14  &  -0.10 $\pm$ 0.00  &  -3.52 $\pm$ 0.30  \\
4XMM J034112.0-050914  &  ASASSN-V J034112.02-050914.5  &  K3.5V  &  5.12  &  12.08  &  2.58  &  13.12  &  1.77E-13  &  -0.58 $\pm$ 0.03  &  -3.48 $\pm$ 0.16  \\
4XMM J034351.2+321309  &  ASASSN-V J034351.24+321309.2  &  K6.5V  &  0.79  &  12.98  &  3.28  &  33.93  &  3.58E-13  &  -1.56 $\pm$ 0.08  &  -2.99 $\pm$ 0.12  \\
4XMM J034425.5+321229  &  ASASSN-V J034425.57+321230.0  &  M3.5V  &  8.38  &  16.19  &  4.85  &  33.27  &  3.94E-14  &  -0.97 $\pm$ 0.05  &  -3.20 $\pm$ 0.25  \\
4XMM J034703.6+240935  &  ASASSN-V J034703.58+240935.0  &  K7V  &  0.30  &  14.44  &  3.32  &  13.44  &  9.06E-14  &  -1.99 $\pm$ 0.10  &  -3.03 $\pm$ 0.10  \\
4XMM J034737.9+232805  &  ASASSN-V J034737.76+232805.2  &  K4.5V  &  8.33  &  13.67  &  2.82  &  13.75  &  5.42E-14  &  -0.41 $\pm$ 0.02  &  -3.42 $\pm$ 0.13  \\
4XMM J034741.3+235818  &  ASASSN-V J034741.44+235819.0  &  K5V  &  0.48  &  13.50  &  2.97  &  13.55  &  1.16E-13  &  -1.68 $\pm$ 0.08  &  -3.19 $\pm$ 0.13  \\
4XMM J034810.9+233025  &  ASASSN-V J034810.98+233025.3  &  K5.5V  &  7.69  &  14.36  &  3.03  &  13.57  &  6.35E-14  &  -0.51 $\pm$ 0.03  &  -3.14 $\pm$ 0.12  \\
4XMM J035017.8-502950  &  ASASSN-V J035017.98-502949.8  &  K6.5V  &  0.42  &  14.42  &  3.26  &  25.28  &  6.10E-14  &  -1.84 $\pm$ 0.09  &  -3.17 $\pm$ 0.34  \\
4XMM J042208.1+191520  &  ASASSN-V J042208.27+191521.8  &  M2V  &  5.12  &  12.91  &  4.22  &  3.58  &  8.49E-13  &  -0.99 $\pm$ 0.05  &  -2.96 $\pm$ 0.06  \\
4XMM J042318.2-275912  &  ASASSN-V J042318.14-275910.3  &  K6.5V  &  1.26  &  14.35  &  3.26  &  174.85  &  4.87E-14  &  -1.36 $\pm$ 0.07  &  -3.24 $\pm$ 2.53  \\
4XMM J042923.2-030146  &  ASASSN-V J042923.22-030146.4  &  G8V  &  10.47  &  13.67  &  1.76  &  140.62  &  4.46E-14  &  -0.06 $\pm$ 0.00  &  -2.78 $\pm$ 0.22  \\
4XMM J044056.3-531412  &  ASASSN-V J044056.45-531413.1  &  G9V  &  5.92  &  13.48  &  1.86  &  168.96  &  4.39E-14  &  -0.33 $\pm$ 0.02  &  -3.38 $\pm$ 0.70  \\
4XMM J045635.3+543505  &  ASASSN-V J045635.17+543506.1  &  K5.5V  &  0.69  &  15.13  &  3.02  &  33.53  &  5.01E-14  &  -1.56 $\pm$ 0.08  &  -2.93 $\pm$ 0.29  \\
4XMM J052951.5+114030  &  ASASSN-V J052951.62+114031.6  &  M1V  &  3.24  &  14.87  &  3.97  &  47.21  &  1.59E-13  &  -1.13 $\pm$ 0.06  &  -2.77 $\pm$ 0.26  \\
4XMM J053101.4+103332  &  ASASSN-V J053101.40+103331.9  &  K3.5V  &  5.16  &  13.88  &  2.55  &  39.00  &  1.08E-13  &  -0.57 $\pm$ 0.03  &  -2.95 $\pm$ 0.22  \\
4XMM J053338.5-242306  &  ASASSN-V J053338.63-242305.1  &  G8V  &  4.03  &  13.74  &  1.73  &  51.91  &  8.77E-15  &  -0.48 $\pm$ 0.02  &  -3.97 $\pm$ 0.83  \\
4XMM J053403.0-053657  &  ASASSN-V J053403.00-053657.3  &  M3V  &  2.65  &  13.69  &  4.56  &  39.96  &  2.49E-13  &  -1.37 $\pm$ 0.07  &  -3.21 $\pm$ 0.08  \\

\hline\hline
\end{tabular}}

\label{table_asas} 

\end{sidewaystable*}

\end{document}